\documentstyle[12pt]{article}
\def\dfrac{\displaystyle\frac}
\def\i{\imath}
\def\a{\alpha}
\def\sn{{\rm sign}}

\def\d{\partial}

\def\noi{\noindent}
\def\ve{\varepsilon}

\newcommand{\bea}{\begin{eqnarray}}
\newcommand{\eea}{\end{eqnarray}}
\newcommand{\be}{\begin{equation}}
\newcommand{\ee}{\end{equation}}
\newcommand{\bc}{\begin{center}}
\newcommand{\ec}{\end{center}}
\newcommand{\ba}{\begin{array}}
\newcommand{\ea}{\end{array}}

\newcommand{\ijmp}[3]{{\it Int. J. Mod. Phys. } {{\bf #1} {(#2)} {#3}}}

\newcommand{\pl}[3]{{\it  Phys. Lett. }{{\bf #1} {(#2)} {#3}}}
\newcommand{\rmp}[3]{{\it  Rev. Mod. Phys.} {{\bf #1} {(#2)} {#3}}}

\normalsize

\begin{document}

\thispagestyle{empty}
\begin{flushright}                              FIAN/TD/96-27\\
                                                hep-th/9612225\\
                                                December 1996

\vspace{0.5cm}
\end{flushright}
\bc
\normalsize
{\large\bf Comment on

"Dynamical Chern-Simons term generation at finite
density"

and "Chern-Simons term at finite density"}

\vspace{2ex}

{\large Vadim Zeitlin\footnote{E-mail address: zeitlin@lpi.ac.ru}}

Department of Theoretical Physics, P.~N.~Lebedev Physical
Institute,

Leninsky prospect 53, 117924 Moscow, Russia

\vspace{4ex}

\ec

\centerline{{\large\bf Abstract}}

\normalsize
\begin{quote}
We comment on the calculation of the Chern-Simons coefficient in
(2+1)-dimensional gauge theories at finite chemical potential made by
A.N.Sissakian, O.Yu.Shevchenko and S.B.Solganik (hep-th/9608159   and
hep-th/9612140).
\end{quote}

\bigskip
In the recent papers A.N.Sissakian, O.Yu.Shevchenko and
S.B. Solganik have considered induced Chern-Simons
coefficient (a coefficient in front of $\i\ve_{\mu\nu\a}p^\a$ tensor structure
in the polarization operator $\Pi_{\mu\nu}(p)$ in the limit $p\rightarrow 0$)
in the (2+1)-dimensional gauge theories at finite chemical potential $\mu$ with
and without constant magnetic field $B$ [1,2].  However, results obtained in
Refs.  [1,2] contradict previous calculations of the Chern-Simons coefficient
(CSC) [3-6] and we cannot accept the former. We shall illustrate our objections
for QED$_{2+1}$.

In Ref. [1] CSC was obtained by calculating  the induced charge at
$B,\mu \ne 0$. The Chern-Simons coefficient was found to be independent of
the magnetic field and equal to (Eq. (12)):

$$
J^{cs} = \frac{e^2}{4\pi} \sn(m) [1- \sn (\mu) \theta(\mu^2 -m^2)]\quad.
\eqno{(1)}
$$

This does not agree with results of Refs. [3-6], where CSC was calculated both
directly and via induced current. The explanation of the difference is the
following: when deriving Eq. (1)  the authors suggested that
only {\it part} of the induced charge (Ref. [1],  Eq. (10) and above)
contributes to CSC, which fact is not correct. The Chern-Simons coefficient is
equal to [3]

$$
J^{cs}(B,\mu) = \dfrac{\d j_0(B,\mu)}{\d B}\quad,
\eqno{(2)}
$$

\noi
and the part omitted in Ref. [1] is $B$-dependent. With the complete expression
for the induced charge the result for $B,\mu,m >0$ is the following [3,6]

        $$
        J^{cs}(B,\mu) = \frac{e^2}{4\pi} +
        \left\{
        \begin{array}{cl}
        {}~~\dfrac{e^2}{2\pi}
        \left[
        \dfrac{\mu^2 -m^2}{2eB} \right]
        ,\quad     & \mu>\/m;
        \\
        &\\
        0,                   \quad             & |\mu|<m;\\
        &\\
        {}-
        \dfrac{e^2}{2\pi}
        \left(
        \left[
        \dfrac{\mu^2 - m^2}{2eB}      \right]
        + 1      \right) ,\quad &\mu<- m,
        \end{array}\right.\nonumber
        \eqno{(3)}
        $$

\noi
where $[ \dots ]$ denotes the integral part.

\bigskip
In Ref. [2] CSC at $B=0, \mu \ne 0$ was obtained to be the same as in Eq. (1)
by calculating the one-loop polarization operator with the external
momentum taken to be zero from the very beginning (Eqs. (4) -- (6), [2]).
However, the momentum integration should be
carried out carefully, since the integrand has singularities, and the limits
$p_0 = 0, {\bf p}^2 \rightarrow 0$ and ${\bf p}^2=0, p_0 \rightarrow 0$ at
$|\mu| > |m|$ are not the same [7].  After making the relevant calculations in
the static limit $p_0 = 0, {\bf p}^2 \rightarrow 0$ one finally obtains the
following expression for the Chern-Simons coefficient:

$$
J^{cs}(\mu) =
\frac{e^2}{4\pi}\sn (m)
(1-\theta(\mu^2-m^2))
\eqno{(4)}
$$

\noi
i.e. the induced CSC at $B,T=0$ vanishes for $|\mu|>|m|$. This is in
agreement with Ref. [8] (but we disagree with  calculations of CSC at $B\ne0$
therein).

\bigskip
This work was supported in part by
RBRF grants $N^o$ 96-02-16210-a and 96-02-16287-a.

\end{document}